\documentclass[english,aip,jcp]{revtex4-1}
\usepackage[T1]{fontenc}
\usepackage[latin9]{inputenc}
\usepackage{amstext}
\usepackage{amssymb}
\usepackage{graphicx}

\makeatletter

\providecommand{\tabularnewline}{\\}

\@ifundefined{textcolor}{}
{%
 \definecolor{BLACK}{gray}{0}
 \definecolor{WHITE}{gray}{1}
 \definecolor{RED}{rgb}{1,0,0}
 \definecolor{GREEN}{rgb}{0,1,0}
 \definecolor{BLUE}{rgb}{0,0,1}
 \definecolor{CYAN}{cmyk}{1,0,0,0}
 \definecolor{MAGENTA}{cmyk}{0,1,0,0}
 \definecolor{YELLOW}{cmyk}{0,0,1,0}
 }

\makeatother

\usepackage{babel}
\begin{document}

\title{First Principles Study of Adsorption of $O_{2}$ on Al Surface with
Hybrid Functionals }

\author{Heng-Rui Liu, Hongjun Xiang and X. G. Gong}

\affiliation{Key Laboratory for Computational Physical Sciences (MOE) and Surface
Physics Laboratory, Fudan University, Shanghai-200433, P. R. China}
\begin{abstract}
Adsorption of $O_{2}$ molecule on Al surface has been a long standing
puzzle for the first principles calculation. We have studied the adsorption
of $O_{2}$ molecule on the Al(111) surface using hybrid functionals.
In contrast to the previous LDA/GGA, the present calculations with
hybrid functionals successfully predict that $O_{2}$ molecule can
be absorbed on the Al(111) surface with a barrier around 0.2$\thicksim$0.4
eV , which is in good agreement with experiments. Our calculations
predict that the LUMO of $O_{2}$ molecule is higher than the Fermi
level of the Al(111) surface, which is responsible for the barrier
of the $O_{2}$ adsorption.
\end{abstract}


\maketitle

\section{INTRODUCTION}

The adsorption and dissociation of oxygen molecule on metal surfaces
is a phenomenon of large relevance to materials processing such as
catalysis and corrosion\cite{background}. Because of the absence
of d-electrons and simple geometric structure, the $O_{2}/Al(111)$
system is considerably simpler than transition metals and thus generally
proposed as a prototype system for oxidation. In order to gain some
insights into the microscopic mechanism of the oxidation of Al(111)
surface, extensive experimental\cite{exp,exp-1,exp-2,exp-3,exp-4}
and theoretical\cite{a-dft,a-dft-1,a-dft-2,a-dft-3,na-dft,na-dft-1,na-dft-2,na-dft-3,na-dft-4}
studies had been done during the past several decades. 

However, even for such a simple $O_{2}/Al(111)$ system, dramatic
disagreement lies between experiments and up-to-date theoretical results
based on the density functional theory (DFT) calculations. Experimentally,
independent molecule beam scattering experiments had consistently
shown\cite{exp,exp-1} that the initial sticking probability of oxygen
molecule on Al(111) surface, defined as the ratio of sticking events
to the total number of molecule-surface collisions, is small ($\backsimeq$$10^{-2}$
) at low incident translation energy, which indicates a barrier around
0.3 eV along the adsorption trajectory. On the other hand, according
to the adiabatic potential energy surface (PES) obtained from the
DFT calculations\cite{a-dft,a-dft-1,a-dft-2,a-dft-3}, the adsorption
of oxygen molecule on the Al(111) surface is exothermic and no sizable
barriers was found.

It seems to be widely established that the failure of DFT to reproduce
the experimental results is due to the adiabatic approximation, which
plays a fundamental role in the frame of DFT. From this point of view,
during the initial adsorption the charge transfer from the Al(111)
surface to $O_{2}$ molecule would be forbidden by Wigner's spin-selection
rules\cite{spin-selection}, and as a result the $O_{2}/Al(111)$
system is confined to some kind of excited states, the description
of which is beyond the scope of the DFT. Many models designed to account
for the non-adiabatic effects had been proposed and barriers can indeed
be produced along the initial adsorption trajectories\cite{na-dft,na-dft-1,na-dft-2,na-dft-3,na-dft-4}.
To invoke all these methods, one has to abandon the single PES picture
and to at least partially abandon the DFT, the cost of which is thus
not very satisfactory and has been under debate\cite{debate,debate-1}.

The other possible reason for the lack of barriers in the previous
DFT calculations, which appears to be overlooked, is that the accuracy
of the exchange-correlation (XC) functionals used before are not good
enough to correctly describe the $O_{2}/Al(111)$ system. Although
DFT methods have impressively proven their computational relevance
during the last decades and won great success in the description of
various systems, the frequently used local-density approximation (LDA)
and gradient-corrected approximation (GGA) functionals have several
severe shortcomings\cite{shortcoming-dft,shortcoming-dft-1,shortcoming-dft-2}.
For example, the calculated formation energy and reaction energy between
small molecules exhibit significant deviations from experimental results.
Considering the structures, the LDA tends to underestimate the bond
length while the GGA often yields a too large value. Moreover, LDA
and GGA always severely underestimate the band gap for insulators
and semiconductors. The HOMO-LUMO gap for small molecules may also
be underestimated in these functionals. For the reaction between oxygen
molecule and aluminum clusters, it has been shown\cite{cluster,cluster-1}
recently that the presently widely used GGA functionals are not adequacy
enough. It has also been shown that for the $O_{2}/Al_{13}^{-}$ case,
the situation can be improved by using hybrid density functionals\cite{cluster}. 

In hybrid functionals the local- or semi-local (gradient-corrected)
functional is modified by adding a fraction of the exact exchange
energy. This represents a sensible compromise between the two simple
mean-field methods {[}DFT and Hartree-Fock (HF){]} and often, probably
due to its partial correction of the self-interaction error in local
and semi-local DFT functionals, yields fairly good agreements with
state of the art many-body methods such as configuration interaction
and perturbation theory as well as with experimental results. The
formation energy and geometries of small molecules, for instance,
are in better agreement with experiments compared to LDA and GGA,
and even the prediction of the band gap and band structure can also
be improved\cite{hybrid-success,hybrid-success-1,hybrid-compare-realize}.

In this paper, the initial adsorption of oxygen molecule on Al(111)
surface is studied by three different functionals, one is the semi-local
Perdew-Burke-Ernzerhof (PBE) functional and the other two are hybrid
functionals (HSE06 and PBE0). The PBE calculation shows a barrier-less
reaction path while both of the two hybrid functionals produce barriers
along the adsorption trajectories, with the barrier height in good
agreement with experimental results. The relative position of the
Fermi surface of the Al(111) surface and the LUMO level of oxygen
molecule is shown to play an essential role in the existence of such
barriers.

\section{THE CALCULATION METHOD}

The calculations presented in this paper are based on the DFT and
performed by means of the plane-wave-pseudoptential code Vienna ab-initio
simulation package (VASP)\cite{vasp-1,vasp-2,vasp-3}. The wave functions
are expanded in a plane-wave basis set, and pseudoptentials are used
to describe the electron-ion interactions. The cutoff kinetic energy
for the plane-wave basis is chosen to be 400 eV for all the calculations.
The geometric structures of the ions are optimized by the conjugate
gradient method. To obtain the electronic minimization, all band simultaneous
update of wave functions is selected as the algorithm for the oxygen
molecule, and a damped velocity friction algorithm is used for the
calculations on the Al(111) surface and the $O_{2}/Al(111)$ system.
The occupation numbers are updated by the Methfessel-Paxton scheme
and a finite electronic temperature of 0.1 eV is used. All total energies
are then extrapolated to zero electronic temperature. The semi-local
Perdew-Burke-Ernzerhof (PBE) functional is used as the basic exchange-correlation
(XC) functional in all the calculations. The hybrid calculations are
done with PBE0 and HSE06 functionals\cite{hse-pbe0,hse-pbe0-1,hse-pbe0-2}.

The calculations on the $O_{2}/Al(111)$ system are restricted to
triplet state in order to account for the open shell nature of the
$O_{2}$ molecule\cite{o2+Si(111)}. The Al(111) surface is modeled
by a seven layer thick slab with a 2$\times$2 surface cell. The vacuum
between slabs is 23 $\textrm{\AA}$, which is large enough to make
the interaction between different slabs negligible. The calculations
for Al(111) surface and $O_{2}/Al(111)$ systems are carried out using
a 4$\times$4$\times$1 Monkhorst-Pack (MP) grid of special k-points.
To check the convergence with respect to the number of k-points, the
calculations for some special points along the reaction paths such
as the positions of the barriers are redone using a 10$\times$10$\times$1
MP grid of k-points. The differences between the adsorption energies
obtained by the two grids are within just a few meV. and thus negligible
compared to the adsorption energies, which are typically greater than
100 meV. To map out a reaction path, we have optimized the structures
for $O_{2}$ molecule with fixed distance to the surface between 1.8
and 3.0 $\textrm{\AA}$, while at each step the two oxygen atoms and
top three layers of Al atoms are relaxed by the PBE functional. For
hybrid calculations the total energy are obtained from static calculations
on the structures obtained by the PBE functional.

\section{RESULTS AND CONCLUSIONS}

\subsection{Free $O_{2}$ molecule}

The ground state of free $O_{2}$ molecule is experimentally found
to be a spin triplet state, and the equilibrium bond length and the
binding energy are 1.21 $\textrm{\AA}$ and 5.2 eV\cite{exp-oxygen},
respectively. The theoretical findings by different functionals along
with the experimental results are summarized in Table 1. The results
are obtained by spin-unrestricted calculations and all of three functionals
can correctly predict the spin triplet state. The results by both
of two hybrid functionals show significant improvement over the PBE
functional especially for the prediction of binding energies and even
the bond length. Our findings are consistent with previous works\cite{hybrid-compare-realize}.

To get more information about the electronic structure of the oxygen
molecule, the density of states (DOS) of free $O_{2}$ molecule are
calculated by the three different functionals and the results are
plotted in Figure 1. The energy level of the occupied orbitals of
free oxygen molecule produced by these three functionals are similar.
The most significant difference between the PBE and the hybrid functionals,
which can be obviously seen from Figure 1, is the gap between the
highest occupied molecular orbitals (HOMO) and the lowest unoccupied
molecular orbitals (LUMO). The precise values of the HOMO-LUMO gap
are 2.27, 5.28 and 6.06 eV for PBE, HSE06 and PBE0, respectively.
As we will see below, the energy levels of the molecular orbitals
in oxygen molecule, especially that of the LUMO, play an important
role in the initial adsorption for $O_{2}$ on the Al(111) surface.

\subsection{Adsorption of oxygen molecule on Al(111) surface}

With the PBE functional we have explored four starting geometries,
and two distinct type of trajectories are found, the final configurations
of which are given in Figure 2. Our results are consistent with previous
studies\cite{a-dft,a-dft-3}.

Two distinct adsorption trajectories by three different functionals
are plotted in Figure 3. Two paths labeled by top and bridge correspond
to those in Figure 2, respectively. The reaction coordinate is the
height from Al(111) surface to oxygen molecule, oxygen molecule is
parallel to the surface and the O-O bond length is optimized at each
height by the PBE functional. As has been mentioned before, in all
these calculations the $O_{2}/Al(111)$ system is constrained to triplet
state to account for the open shell nature of the oxygen molecule.
The adsorption energy is defined as:
\[
E_{ad}=E_{O_{2}/Al(111)}-E_{O_{2}}-E_{Al(111)}
\]
where E denotes the total energy obtained by the DFT.

It turns out that in the PBE calculation the adsorption energy decreases
monotonically along the reaction coordinate for both trajectories,
which indicates a non-activated reaction process. On the other hand,
both hybrid calculations produce barriers around 0.2$\sim$0.4 eV
along two trajectories, which are consistent with experimental results\cite{exp,exp-1}.

We do the Bader charge population analysis\cite{Bader,Bader-1,Bader-2}
along the two trajectories by the three functionals and the results
along the bridge adsorption path is plotted in Figure 4, the charge
along the top adsorption path is qualitatively the same. It can be
clearly seen from this figure that the charge is transferred from
the Al(111) surface to the molecule, which is qualitatively consistent
with previous findings\cite{a-dft}. Moreover, at large distance (the
height $>2.4\textrm{ \AA}$) PBE predicts more charge transfer than
that of the hybrid functionals. This difference of charge transfer
between the PBE and hybrid functionals could account for the different
predictions of the adsorption energies and thus the existence of barriers,
similar to the $O_{2}/Al_{13}^{-}$ case\cite{cluster}.

In order to understand why the hybrid functional can predict a adsorption
barrier, We calculate the LUMO of free oxygen molecule and the Fermi
level $E_{F}$ of the clean Al(111) surface by different functionals
and the results are illustrated in Figure 5. In order to compare the
energy levels from different systems, the presented values of the
energy levels are relative to the energy levels of vacuum. It can
be seen from Figure 5 that the Fermi level of the clean Al(111) surface
is nearly the same by the three different functionals. The LUMO of
oxygen molecule lies below the $E_{F}$ of the Al(111) surface by
the PBE functional, allowing a spontaneous charge transfer to the
LUMO of the molecule when it approaches the surface. Since the LUMO
is an anti-bonding orbital, the filling of such orbital drives the
dissociation of $O_{2}$. In contrast with the result of the PBE,
the LUMO of oxygen molecule is above the $E_{F}$ of Al(111) surface
when using hybrid functionals, which gives rise to a barrier in charge
transfer to the LUMO of $O_{2}$ molecule and leads to the activated
adsorption process. Thus the results in Figure 5 are consistent with
that in Figure 4 and finally lead to the different predictions of
the adsorption paths in Figure 3 by the three functionals. It would
be interesting to note that for the $CO/Pt(111)$ system, the position
of the energy level of the LUMO of the CO molecule was reported to
play an key role in the prediction of the preference adsorption site\cite{CO-Pt(111)}.

In conclusion, we have presented DFT total-energy calculations for
the adsorption of the oxygen molecule on the Al(111) surface using
different exchange-correlation functionals. In contrast to semi-local
functional, hybrid functionals predict a barrier for the adsorption
of $O_{2}$ on Al(111) surface, in agreement with experiments. Our
results suggest that the LUMO of $O_{2}$ molecule is higher than
the Fermi level of the Al(111) surface, which is responsible for the
barrier of the $O_{2}$ adsorption. To the best of our knowledge,
it is the first time that the barriers have been successfully produced
for this reaction within the scope of the DFT. We hope that our findings
can offer a new perspective for the understanding of the microscopic
mechanism of the oxidation of the Al(111) surface.

This work is partially supported by the Special Funds for Major State
Basic Research , National Science Foundation of China , Ministry of
Education and Shanghai Municipality. The calculations were performed
in the Supercomputer Center of Shanghai, the Supercomputer Center
of Fudan University.

\clearpage{}
\begin{table}[p]
\noindent \begin{raggedright}
\caption{The equilibrium spin, bond length and binding energy for free $O_{2}$
molecule by different functionals along with experimental results.
The calculation results by two hybrid functionals are in good agreement
with experiment\cite{exp-oxygen}. }

\par\end{raggedright}

\begin{ruledtabular} %
\begin{tabular}{cccc}
 & Spin & Bond length ($\textrm{\AA}$$\AA$) & Binding energy (eV)\tabularnewline
\hline 
PBE & 1 & 1.23 & 6.67\tabularnewline
HSE06 & 1 & 1.21 & 5.18\tabularnewline
PBE0 & 1 & 1.21 & 5.17\tabularnewline
Experiment & 1 & 1.21 & 5.20\tabularnewline
\end{tabular}\end{ruledtabular}
\end{table}

\clearpage{}

\begin{figure}[p]
\noindent \begin{centering}
\includegraphics[scale=0.5]{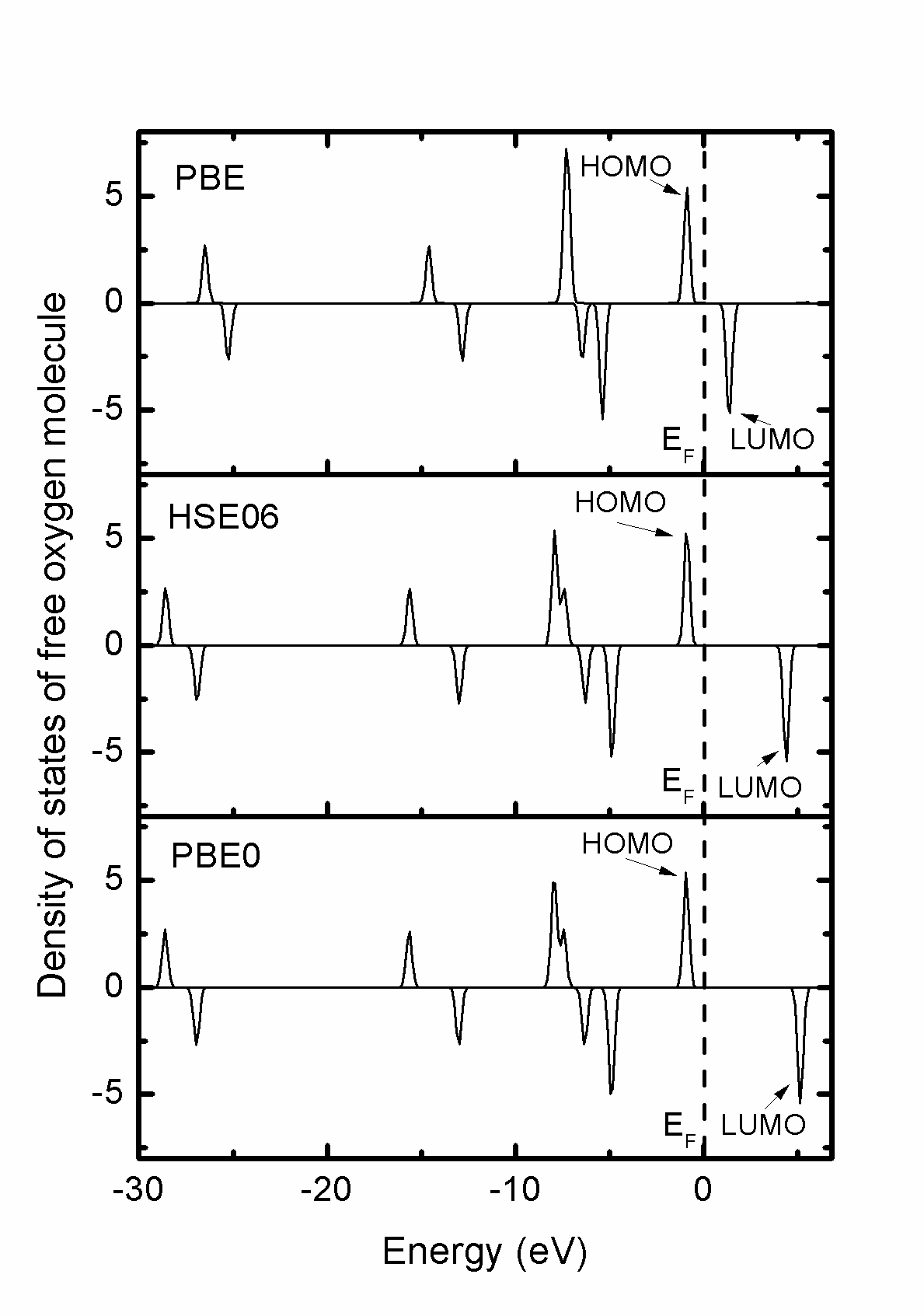}
\par\end{centering}

\caption{Energy levels of the valence orbitals of free $O_{2}$ molecule calculated
by three different functionals. The Fermi energy $E_{F}$ is set to
zero. The most obvious difference between the PBE and hybrid functionals
is the HOMO-LUMO gap. Both of two hybrid functionals show a significant
larger gap than the PBE functional.}
\end{figure}

\clearpage{}

\begin{figure}
\begin{centering}
\includegraphics[scale=0.55]{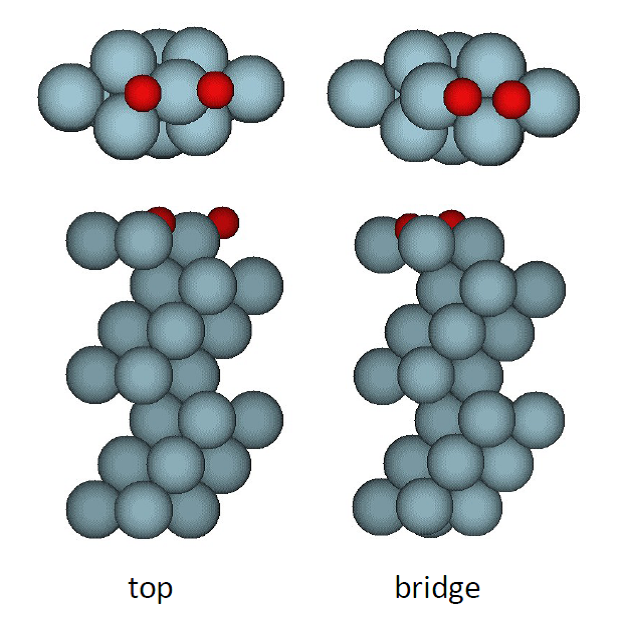}
\par\end{centering}

\caption{Two different adsorption geometries obtained by the PBE calculation.
Oxygen molecule is parallel to the surface and dissociates in both
trajectories and the O-O distance is 3.3 $\textrm{\AA}$ in the top
adsorption and 2.2 $\textrm{\AA}$ in the bridge adsorption. The top
adsorption is energetically more favorable and the equilibrium height
is around 0.9 $\textrm{\AA}$, which is in consistent with previous
studies and slightly higher than experimental value (0.6$\sim$0.8
$\textrm{\AA}$)\cite{a-dft,a-dft-3}. }
\end{figure}

\clearpage{}

\begin{figure}
\begin{centering}
\includegraphics[scale=0.5]{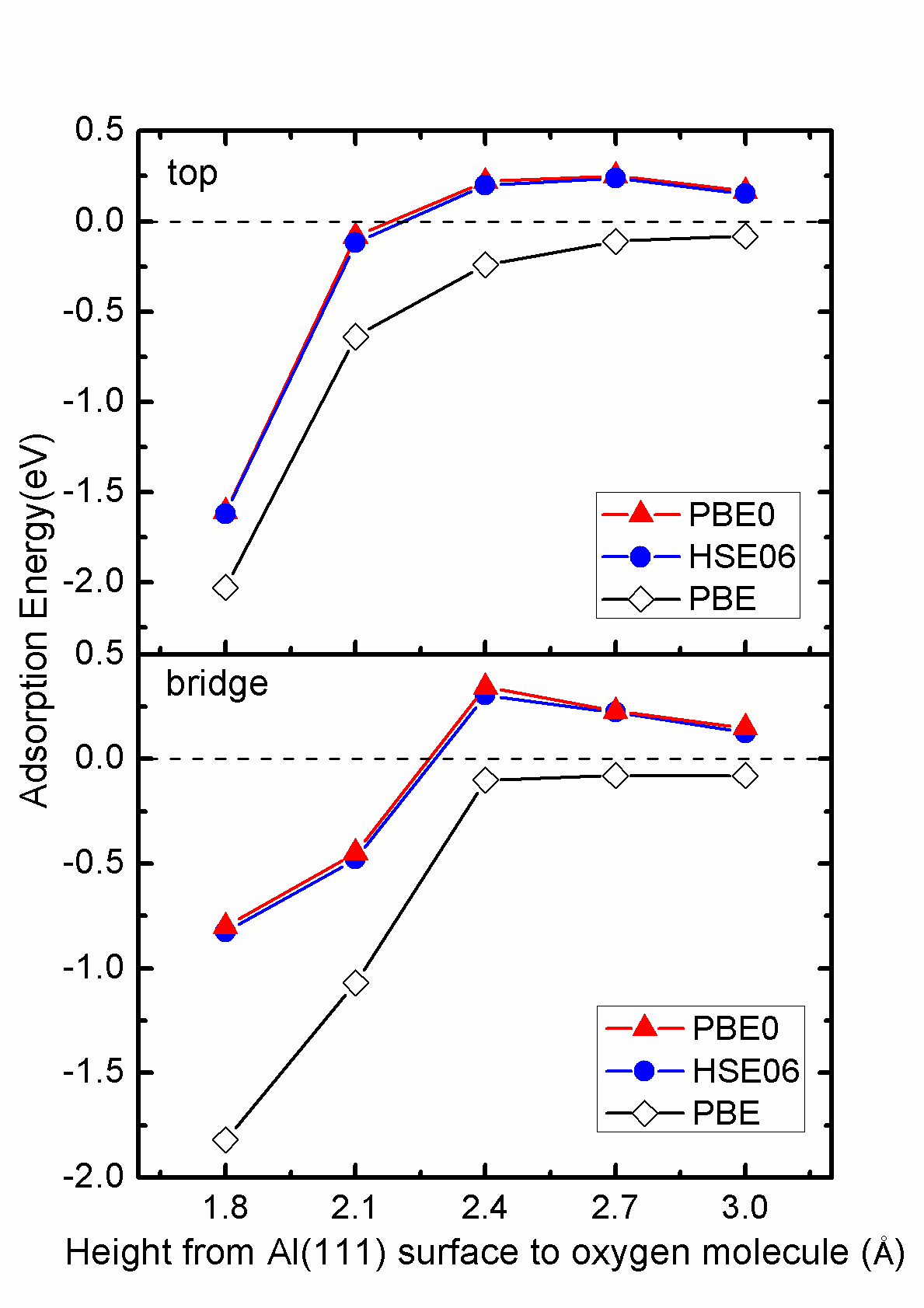}
\par\end{centering}

\caption{Adsorption energy along two trajectories calculated by different exchange-correlation
functionals. The two paths labeled by top and bridge correspond to
those in Figure 2, respectively. In the PBE calculation the adsorption
energy decreases monotonically for both trajectories and no barrier
is found. On the other hand, both hybrid calculations produce barriers
around 0.2$\sim$0.4 eV along two trajectories, which are consistent
with experimental results.}
\end{figure}

\clearpage{}

\begin{figure}
\begin{centering}
\includegraphics[scale=0.5]{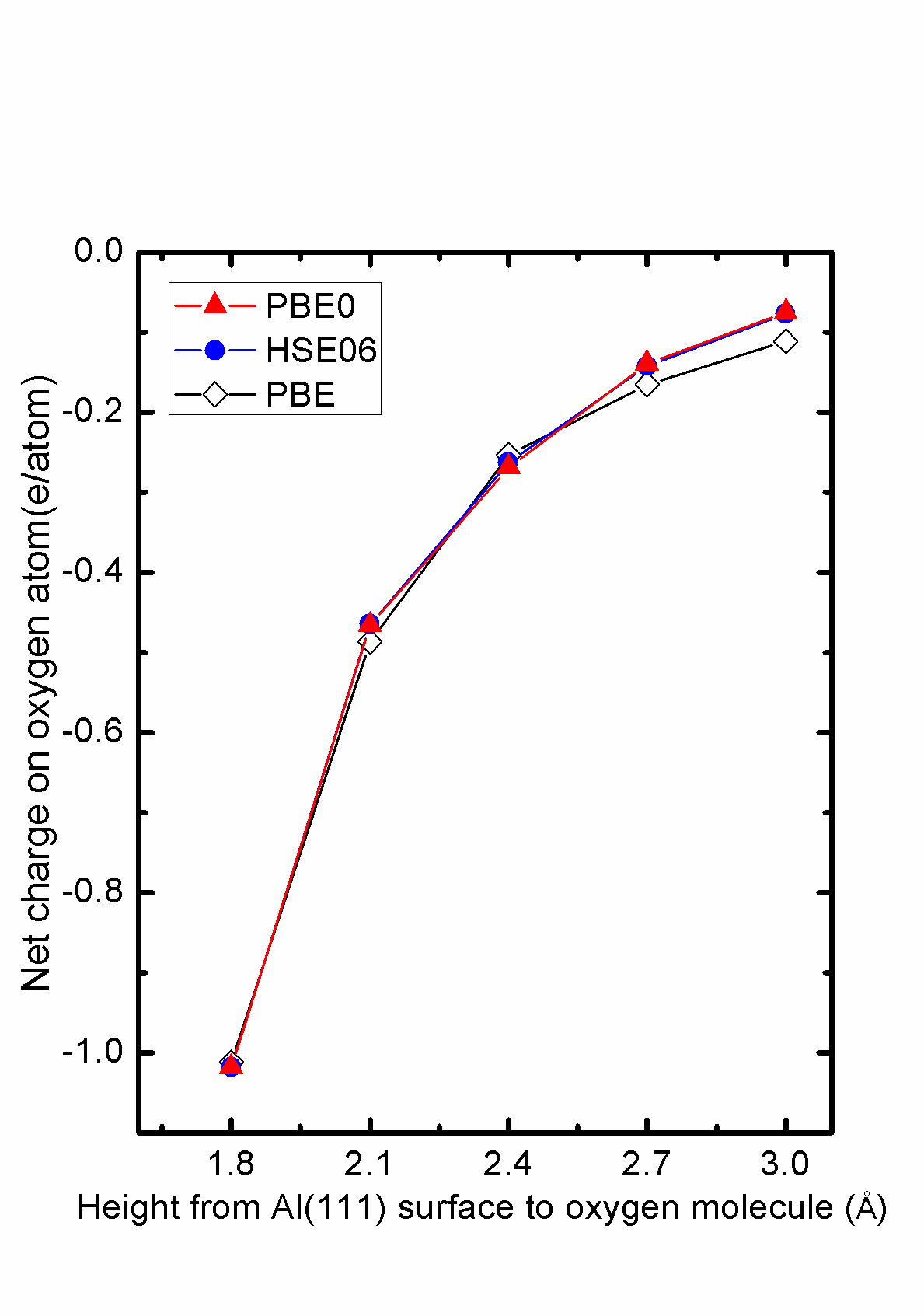}
\par\end{centering}

\caption{Net charge on oxygen atom along the bridge adsorption path for three
different functionals. It can be clearly seen that the charge is transferred
from the Al(111) surface to the molecule. Moreover, at large distance
($>2.4\textrm{ \AA}$), PBE predicts more charge transfer than that
of hybrid functionals.}
\end{figure}

\clearpage{}

\begin{figure}
\begin{centering}
\includegraphics[scale=0.58]{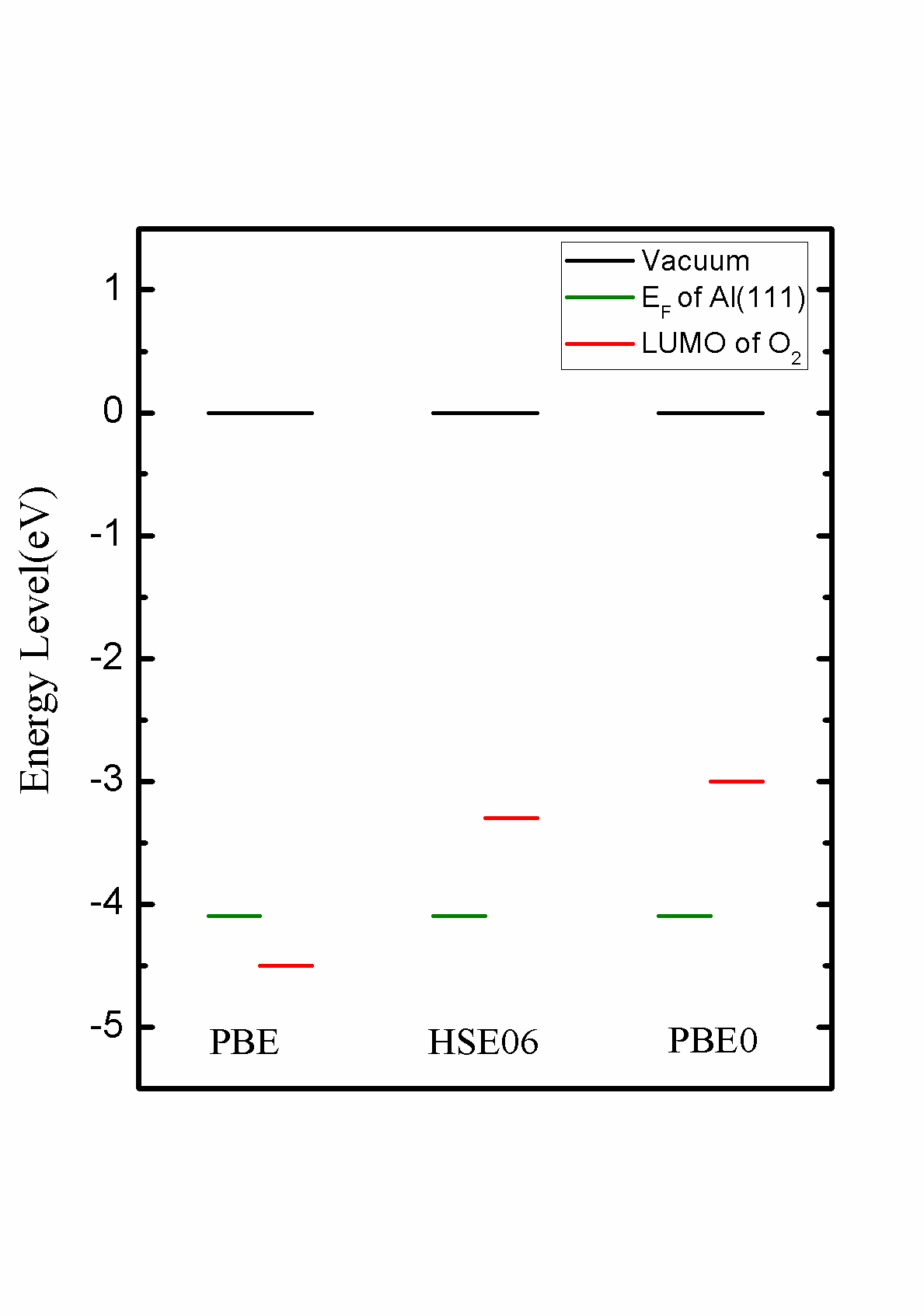}
\par\end{centering}

\caption{The LUMO of free oxygen molecule and the Fermi level $E_{F}$ of the
Al(111) surface calculated by different exchange-correlation functionals.
The energy level of vacuum is set to zero. The $E_{F}$ of the clean
Al(111) surface is nearly the same by three different functionals.
The LUMO of oxygen molecule lies below the $E_{F}$ of the Al(111)
surface by the PBE functional. In contrast with the result of the
PBE, the LUMO of the oxygen molecule is above the $E_{F}$ of the
Al(111) surface when using the HSE06 and the PBE0 functional.}
\end{figure}

\end{document}